\documentclass[superscriptaddress,prl,floatfix,twocolumn,amsmath,amssymb,aps]{revtex4-1}

\usepackage{graphicx}
\usepackage{amsmath}
\usepackage{amssymb}
\usepackage{dcolumn}
\usepackage{color}
\usepackage{multirow}
\usepackage{bm}
\usepackage{subfigure}

\begin{document}
\title{The dynamic nature of high pressure ice VII}
\author{Qi-Jun Ye}
\affiliation{State Key Laboratory for Artificial Microstructure and Mesoscopic Physics, and School of Physics, Peking University, Beijing 100871, P. R. China}
\author{Lin Zhuang}
\affiliation{State Key Laboratory for Artificial Microstructure and Mesoscopic Physics, and School of Physics, Peking University, Beijing 100871, P. R. China}
\author{Xin-Zheng Li}
\email{xzli@pku.edu.cn}
\affiliation{State Key Laboratory for Artificial Microstructure and Mesoscopic Physics, and School of Physics, Peking University, Beijing 100871, P. R. China}
\affiliation{Collaborative Innovation Center of Quantum Matter, Peking University, Beijing 100871, P. R. China}
\date{\today}
\begin{abstract}
    Starting from Shannon's definition of dynamic entropy, we proposed a simple theory to describe the transition between different rare event related dynamic states in condensed matters, and used it to investigate high pressure ice VII.
    Instead of the thermodynamic intensive quantities such as the temperature and pressure, a dynamic intensive quantity named dynamic field is taken as the controlling variable for the transition.
    Based on the dynamic entropy versus dynamic field curve, two dynamic states corresponding to ice VII and dynamic ice VII were discriminated rigorously in a pure dynamic view.
    Their microscopic differences were assigned to the dynamic patterns of proton transfer.
    This study puts a similar dynamical theory used in earlier studies of glass models on a simple and more fundamental basis, which could be applied to describe the dynamic states of realistic and more condensed matter systems.
\end{abstract}

\maketitle

\begin{figure}[t]
    \centering
    \includegraphics[width=\linewidth]{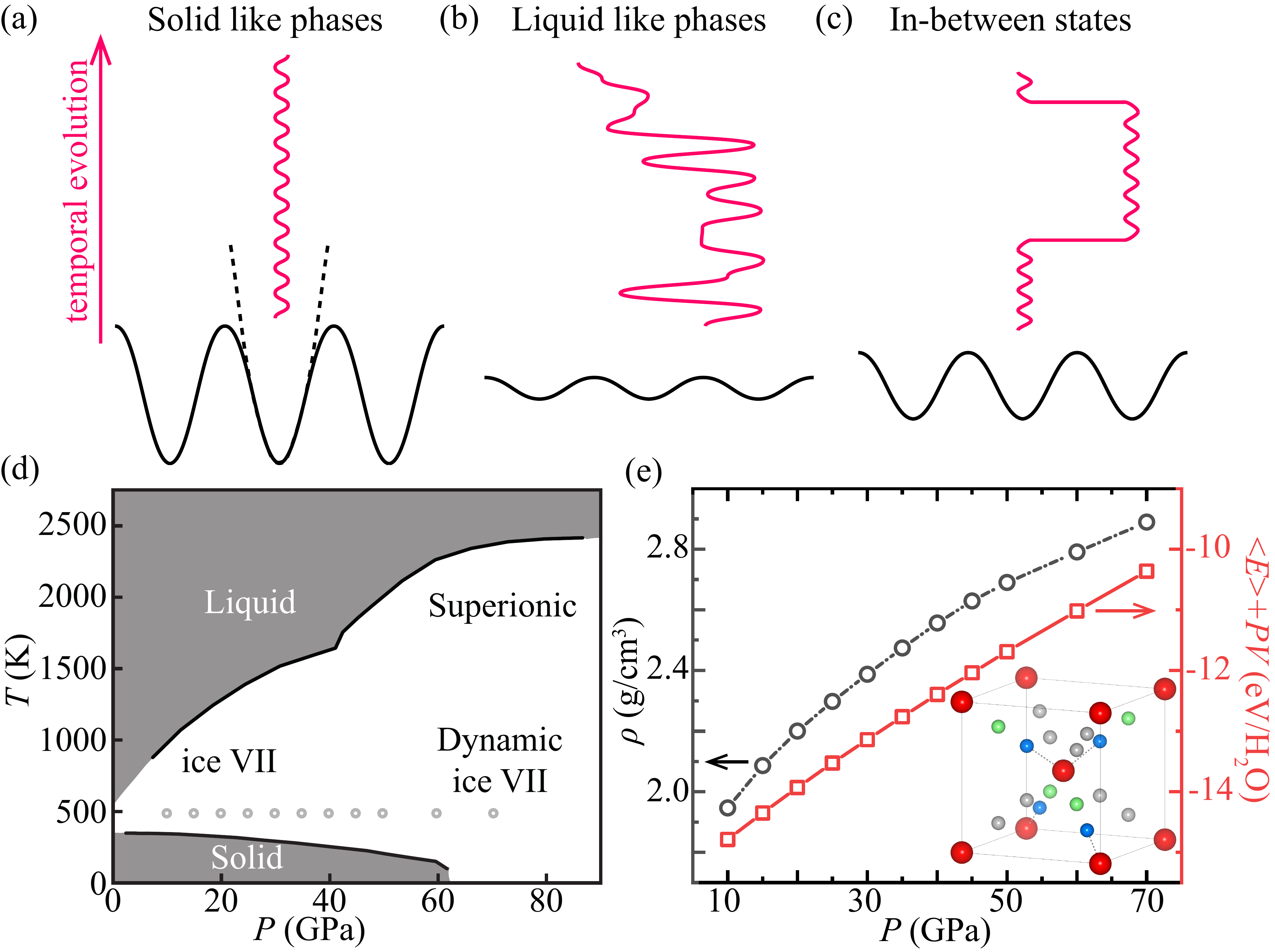}
    \caption{(a)(b) Solid, liquid, and (c) the in-between state. (d) The phase diagram of bcc ice. There are well-defined solid and liquid states in the shadowed region, while in the white region the boundaries between ice VII, dynamic ice VII, and SI remain controversial. (e) Density and thermodynamic state function at 500~K. Inset: bcc ice structure. In the bcc skeleton formed by oxygen (red), protons intersperse in the covalent sites (blue) of neighboring oxygen pairs. The equivalent sites (green) can be occupied through transfer motions. The other sites (grey) consist another hydrogen bonding network.}
    \label{fig1}
\end{figure}

Matters exist in form of states, wherein the physical properties vary continuously before abrupt changes happen upon their transitions~\cite{Landau1937}.
Abundant states have constituted our understanding of matters from different aspects, such as the crystal states characterized by the atomic structures~\cite{Scott1974,Cowley1976}, and the superconductors and charge density waves characterized by the electronic structures~\cite{McMillan1976,Bardeen1957,Gruner1988,Fisher1991}.
Recent years have witnessed considerable progress of theoretical methods on simulating these states, especially the former, along with thorough exploration of the rare events and their dynamic properties~\cite{Peters20171,Hernandez2016}.
Here, rare events mean dynamic activities occurred out of equilibrium and with constraints, e.g. atom A cannot move until atom B move out of the way~\cite{Palmer1984}, and ice rule governs the arrangement of protons~\cite{Bernal1933}.
Description of these dynamic states, constraints, and their transitions using the conventional thermodynamical language of equilibrium states, however, is difficult.
One prominent example of such problems, when rare event related dynamic properties are crucial, exists in high pressure ($P$) water.
At $P$s ranging from 2~GPa to 80~GPa and temperatures ($T$s) of a few hundreds Kelvins, its states of matter are dominated by several similar body-centered-cubic (bcc) structures, e.g. ice VII, dynamically disordered ice VII (dynamic ice VII), and superionic (SI) ice~\cite{Benoit1998,Benoit2002,Goncharov1999,Goncharov2005,Schwegler2008,Bove2013,Guthrie2013,Millot2019,Queyroux2020}.
Conventionally, these states can be attributed to solid with atoms localized in the crystalline sites (Fig.~\ref{fig1}(a)) or liquid with atoms travelling ergodically over the whole configurational space (Fig.~\ref{fig1}(b)).
The so-called dynamic ice VII, however, present an in-between feature, i.e. protons are localized on their sites but can occasionally hop to another ones in a timescale of picoseconds and longer (Fig.~\ref{fig1}(c)).
It was considered as a distinct state from ice VII in earlier studies (Fig.~\ref{fig1}(d)), due to the occurrence of dynamical translational disordering (proton hopping along hydrogen bonds)~\cite{Benoit2002,Goncharov2005}.
One may intuitively interpret this as the protons cannot be transferred in ice VII and can in dynamic ice VII.
However, this criterion is questionable, as the structures of the bcc skeleton of oxygens remain the same and there is no transient change in the structural order or thermodynamic properties (Fig.~\ref{fig1}(e) and in Refs.~[\onlinecite{Hemley1987,Wolanin1997,Loubeyre1999}]).
A paradox arises: if the timescale is long enough, proton transfer can also occur in ice VII.
Consequently, ice VII and dynamic ice VII should be considered as a single state of matter where proton transfer can occur in the long time limit, though with a large numerical variance in the transfer rates.
The answer to this paradox is still absent.
In this article, we put the transition between different bcc ice states in a rigorous footing of dynamical transition.
Starting from Shannon's definition of the dynamic entropy, a simple mathematical form of dynamic
partition function is derived, based on which a dynamical theory is presented.
Dividing the space into pieces of components, the atomic trajectories of protons are decomposed into intra-component vibrational motions and inter-component diffusive motions.
The dynamic field, a central quantity in the dynamic partition function, is taken as the controlling variable for the inter-component motions.
This field can reveal the different patterns of dynamic motions related to dynamic constraint and hence the transitions between dynamic states.
In the simulations, we derived its values for each $P$ by mapping to a constructed referenced system.
Two states were discriminated using the dynamic entropy versus dynamic field curve in the region of static and dynamic ice VII, and a transition $P$ is obtained by approaching the simulation results to the long time limit.
And the mechanism underlying this transition is further detailed.

We shall start with the description of the dynamic properties.
In previous studies, diffusion coefficient was taken as the dynamic order parameter in discriminating static and dynamic ice VII~\cite{Hernandez2016,Hernandez2018,Komatsu2020,Zhuang2020}.
Considering the rare event nature of proton transfer at low $T$s, we resort to a fundamental property, the proton motions.
There are two types:~(1) liquid-like diffusive motion as proton transfers to another equivalent site; (2) solid-like localized motion as proton oscillates around its own equilibrium site.
The essential dynamical information is muffled by thermal noises, since the diffusive motions are rare compared to the localized motions.
In order to highlight the diffusive motions and summarize the localized motions, the concept of ``components'' is introduced~\cite{Palmer1982}.
One component is an ingredient of the phase space, which is composed by a close set of neighboring phase points containing a local minimum of the potential energy surface (PES):
\begin{equation}
    \Omega = \cup_\alpha \Omega_\alpha,\text{ with }\Omega_\alpha = \{(\bm{x}, \bm{p})\in \Omega_\alpha\}.
\end{equation}
$\Omega$ represents the whole phase space and $\Omega_\alpha$ is the component.
The components are assumed to have confinement condition (atoms stay in one component for a long time) and internal ergodicity (equilibrium statistics ensured for intra-component motion)~\cite{Palmer1982}.
They are defined using Voronoi decomposition, by constructing the Wigner-Seitz cell of the equivalent sites of the same kind of atoms~\cite{Wigner1933}.
Upon this, we call each inter-component hopping an activity, and hence define the activity rate $k$ as the number of activities $K$ occurred in a certain observation time $t_{\text{obs}}$, with $k=K/t_{\text{obs}}$.
Extensive dynamic quantities scale with $t_{\text{obs}}$, but when no ambiguity exists, we omit $t_{\text{obs}}$ in the equations.
The core quantity to describe a system is its partition function and the relevant degrees of freedom (DOFs).
Thermodynamic intensive quantities, such as $T$ and $P$, are the acknowledged choice to identify the equilibrium state of a system and hence to reveal how thermodynamic phase transition occurs.
In bcc ice, however, diffusion coefficient presents gradual changes in a wide range of $T$s and $P$s, which are accumulated to qualitative changes, while evident change in structure order and state function cannot be observed.
This implies that an extra DOF other than $T$ and $P$ should be crucial, and a dynamic form of the partition function should be resorted to.
Analogy can be seen in glass transition, where supercooled fluid is prepared via rapid cooling and $T$ does not play a fundamental role~\cite{Debenedetti2001}.
To denote this extra DOF, we use ``dynamic field'', a term proposed in glass transition studies as the intensive quantity~\cite{Whitelam2004}, and the dynamic activity as the extensive quantity.
This field, denoted by $s$, is introduced first as an auxiliary field for enhanced/reduced sampling in analyzing the dynamics of each $(T, P)$~\cite{Merolle2005,Jack2006,Garrahan2007,Hedges2009,Chandler2010,Garrahan2010}.
Interpreting glass transition as a space-time phase transition, Hedges \textit{et al.} proposed that $s$ is its controlling variable~\cite{Hedges2009}.
But this $s$ is a pure predefined mathematical device, so that this scheme is limited to ideal spin or glass former models.
We derive the value of $s$ for each $(T, P)$ by mapping to a referenced systems and hence extend the application of this dynamical theory to realistic bcc ice, as detailed later.
Following the thermodynamic convention, we write the dynamic partition function in form of the sum of probabilities to find system in particular dynamic states, as
\begin{equation}
    \label{partition function}
    Z(s) = \sum_K p(s,K),
\end{equation}
where the partition according to $K$ is applied~\footnote{Here we distinguish the trajectories by the number of activities, not by their specific processes. For example, trajectories $LTL$ and $LLT$ (with $L$ and $T$ represent the local and transfer motion, respectively) belong to the same partition with activity $K=1$. There can be finer partitions for the trajectory space and corresponded partition function and entropy such as the well-known Kolmogorov-Sinai entropy~\cite{Kolmogorov1959}.}.
Eq.~(\ref{partition function}) becomes $Z_0=\sum_K p_0(K)$ when $s=0$, where $p_0(K)$ is the unbiased distribution without artificial samplings.
When assigning finite $s$, $p(s, K)$ takes the form $p_0(K)e^{-s\cdot K}$
by comparing with the thermodynamic formula~\cite{Ruelle1985,Lecomte2007,Hedges2009}.
Here we note that an elegant mathematical form of both the thermodynamic and dynamic partition functions
can be derived using merely the information theory~\cite{Jaynes1957}.
According to Shannon~\cite{Shannon1948}, the dynamic entropy within $t_{\text{obs}}$ is defined as
\begin{equation}
    S_{\text{D}}(s) =-\sum_K p(s, K) \ln p(s, K).
    \label{dynamic entropy}
\end{equation}
A reasonable $p(s,K)$ should give statistical results consistent with the observed ones, i.e. $p(s,K)$ and $\langle K \rangle_s = \sum_K p(s, K) \cdot K$ must conform to $p_0(K)$ and $K_0$.
The variation of $p(s,K)$ subjected to the requirements of the least bias estimation from known results and the maximum of entropy $S_{\text{D}}$~\cite{Jaynes1957} is
\begin{widetext}
    \begin{equation}
        \delta \left\{S_{\text{D}}(s)+\sum_K \lambda_{1,K} \left[p(s,K) - p_0(K)\right] +
        \lambda_2 \left[\langle K \rangle_s - K_0\right]\right\}
        =\sum_K \delta p(s,K) \left\{-(\ln p(s,K) + 1) +\lambda_{1,K} + \lambda_2 K\right\} = 0,
    \end{equation}
\end{widetext}
which leads to
\begin{equation}
    p(s,K) \sim e^{-(\lambda_{1,K} - 1)-\lambda_2 K} \equiv p_0(K) e^{-s\cdot K}.
\end{equation}
Here $\lambda_{1,K}$ and $\lambda_2$ are the Lagrange multipliers of constraint, and we conclude the explicit-$K$ part by defining the field $s=\lambda_2$ and the implicit-$K$ part by $p_0(K)$.
In so doing, the dynamic partition function in form of
\begin{equation}
    \label{partition sum}
    Z(s) = \sum_K e^{-s\cdot K} p_0(K)
\end{equation}
is naturally obtained.
The dynamic field can reveal the intrinsic correspondence among different thermodynamic configurations.
Dynamic properties are fundamentally controlled by $s$, through $\langle K \rangle_s=\frac{\partial Z(s)}{\partial s}$.
Therefore upon artificially changing $s$, Hedges \textit{et al.} claimed that for each specific $(T, P)$ the system would experience a transition between an active state and an inactive one~\cite{Hedges2009}.
Beyond this, we realized that similar $s$-dependencies of $Z(s)$ among different $(T,P)$s can reveal their internal connection.
To quantify this, we resort to a referenced partition function $Z_{\text{ref}}$, which contains all the dynamical information within $(T,P)$ region of the same bcc-ice structure.
In the light of thermodynamics, where $Z_{\text{NVT}}$ is given by the sum of a series of $NVE$ ensembles with weight $e^{-\beta E}$, we construct $Z_{\text{ref}}$ as
\begin{equation}
    Z_{\text{ref}} = \sum_{(T,P)} Z_{0,(T,P)}\cdot e^{-\beta (F+PV)}.
    \label{reference partition sum}
\end{equation}
$Z_{0,(T,P)}$ is the $s=0$ case of Eq.~(\ref{partition sum}), and $e^{-\beta (F+PV)}$ is its weight.
By mapping to $Z_{\text{ref}}$, the referenced dynamic field $s_{\text{ref}}$ is determined for each $(T,P)$.
The rule is to ensure that the expectation value of the referenced $s$-ensemble at $s=s_{\text{ref}}(T,P)$ with $Z_{\text{ref}}$ is the same as its unbiased observation at $s=0$ with $Z_{0,~(T,P)}$, i.e.
\begin{equation}
    \langle A \rangle_{s=0,~Z_{0,(T, P)}} = \langle A \rangle_{s_{\text{ref}}(T,P),~Z_{\text{ref}}}.
    \label{sref}
\end{equation}
This scheme enables learning the transition between different dynamic states solely from $s_{\text{ref}}$.

\begin{figure}[t]
    \centering
    \includegraphics[width=\linewidth]{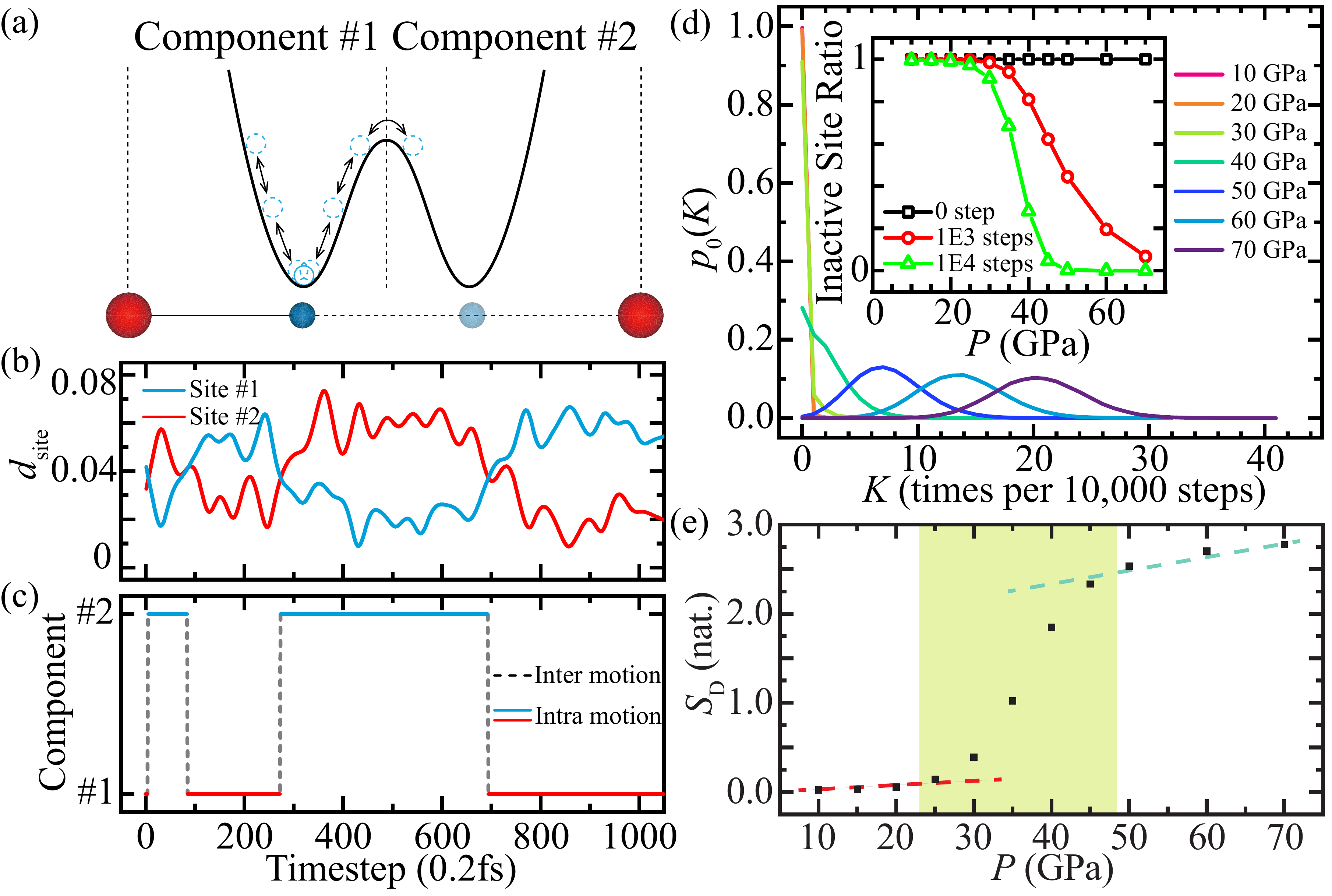}
    \caption{(a) Schematic of the potential energy surface (PES) of proton transfer. (b) the realistic trajectory in coordinate space, and (c) the trajectory in component space. $d_{\text{site}}$ is the distance of the proton to its equilibrium position inside each component, in the unit of fractional coordinate of the simulation cell. (d) $p_0(K)$ at different $P$s with $t_{\text{obs}}=10,000~$steps. The inset of (d) shows the $t_{\text{obs}}$-dependence of the percentage of inactive sites for $K(t)=0$. (e) $S_{\text{D}}$ computed from $p_0(K)$. The dashed lines and shadowed region are guided for the eye.}
    \label{fig2}
\end{figure}
\begin{figure}[t]
    \centering
    \includegraphics[width=\linewidth]{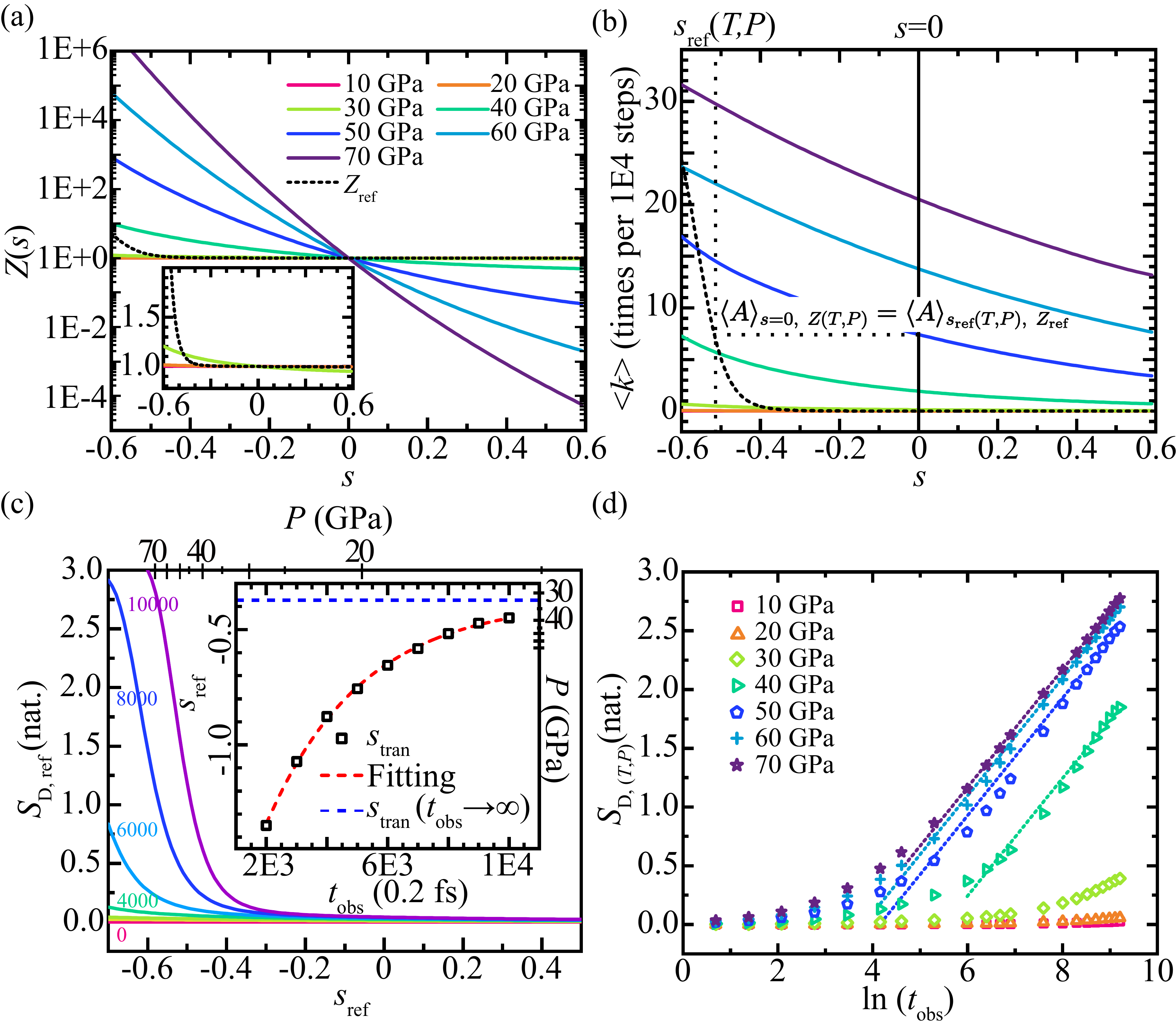}
    \caption{(a) The partition function of the $s$-ensemble and (b) $\langle K \rangle_s$ as its ensemble average at different $P$s and 500K. (a)(b) share the legend. All the data is presented with $t_{\text{obs}}=$2~ps (10,000 steps). (c) The $S_{\text{D,ref}}$ with different $t_{\text{obs}}$ (in the unit of steps). Inset of (c): the tendency of transition point $s_{\text{Tran}}$ towards long time limit. (d) The tendency of $S_{\text{D},(T,P)}$ on $t_{\text{obs}}$. The dashed lines show theoretical results of Eq.~(\ref{long time entropy}). }
    \label{fig3}
\end{figure}

In order to sample the trajectory space, i.e. capture the activities and further derive $Z_{\text{ref}}$ and $s_{\text{ref}}$, we performed extensive molecular dynamic simulations.
This is enabled by resorting to a machine learning potential~\cite{Wang2018,Zhuang2020}, and simulations consisted by samplings up to 1E7 timesteps and with the timescale of a few nanoseconds for each ($T$, $P$).
For more details of these simulations, please see our supplemental material~\cite{SI}.
The protons are judged which component they belong to for each timestep, as shown in Fig.~\ref{fig2}(a)-(c).
The trajectories are decomposed into intra-component localized motions (Fig.~\ref{fig2}(c), vertical solid lines), and inter-component diffusive motions (Fig.~\ref{fig2}(c), dashed lines).
We count the number of activities at each component and present the total distribution $p_0(K)$ in Fig.~\ref{fig2}(d).
At 10-30~GPa, $p_0(K)$ concentrates at $K=0$ and fall sharply at non-zero values, corresponding to the fact that few proton transfer happens during our simulations.
Above 40~GPa, there are finite rate peaks, which extend to higher values with increasing $P$s, indicating easier proton transfers.
Consistent with this trend, the inactive sites ($K=0$) are dominated at low $P$ and gradually drop to zero at high $P$s (the inset of Fig.~\ref{fig2}(d)).
The crucial role of $t_{\text{obs}}$ will be discussed later.
$S_{\text{D}}$ is also computed, telling two distinct regions corresponding to ice VII and dynamic ice VII (Fig.~\ref{fig3}(e)).
However, the gradual transition with wide $P$ range (Fig.~\ref{fig3}(e), shadowed region), which does not disappear with increasing the simulation scale, is beyond the scope of $P$-controlled phase transition (where the transition should be abrupt in $P$).
In other words, $P$ can reveal the dynamic difference, but a native dynamic perspective is more requisite.
One unique perspective in describing such a transition is offered by $s$.
The exponential factor $e^{-s\cdot K}$ in $Z(s)$ play the role of raising the contribution of inactive components and reducing that of active ones.
When $s$ is increased (decreased), the activities are suppressed (enhanced).
This trend applies to all simulated $P$s.
However, there are two different $s$-dependencies for $Z(s)$ and activity rate $\langle k \rangle_s$.
These dynamic quantities are insensitive to the change of $s$ at low $P$s while they can be motivated by lowering $s$ at high $P$s, as shown by the almost vertical lines at low $P$s and the finite slopes at high $P$s in Fig.~\ref{fig3}(a) and \ref{fig3}(b).
This qualitative difference can be more clearly seen via the state function $S_\text{D}$ versus $s_{\text{ref}}$ curve in Fig.~\ref{fig3}(c), where $s_{\text{ref}}$ is defined for each $P$ using Eq.~(\ref{sref}) as shown in Fig.~\ref{fig3}(b).
There are two distinct regions: a large-$s_{\text{ref}}$ region
with steady and nearly zero $S_\text{D}$, mapping to low $P$s; and a small-$s_{\text{ref}}$ region with rapidly increased $S_\text{D}$, mapping to high $P$s (Fig.~\ref{fig3}(c)).
Besides, we notice that the change in slope becomes sharper with increasing $t_{\text{obs}}$.
Due to the limited sampling scale, the transition point is practically determined by the intersection of extrapolated lines of two ends at finite $t_{\text{obs}}$~\cite{SI}.
Its dependencies on $t_{\text{obs}}$ can be well-fitted by an exponential form (inset of Fig.~\ref{fig3}(c)).
The converged value for $t\to\infty$ is $s_{\text{ref}}=-0.37$, which maps to $P\approx 32$GPa.
We note that this $P$ is consistent with Ref.~[\onlinecite{Hernandez2018}].
The behavior of $S_{\text{D}}$ towards long time limit roots in homogeneouity.
When system is dynamically homogeneous, the activities are globally uniformed.
According to the central limit theorem, $p(K,t_{\text{obs}}\to\infty)$ approaches a normal distribution, as
\begin{equation}
    \begin{split}
        p(K,t_{\text{obs}})
        &\sim \mathcal{N}(\overline{K}(t_{\text{obs}})=\overline{k}t_{\text{obs}}, \sigma_{t_{\text{obs}}}^2) \\
        &= \frac{1}{\sqrt{2\pi}\sigma_{t_{\text{obs}}}} \exp \left[-\frac{(K-\overline{k}t_{\text{obs}})^2}{2\sigma_{t_{\text{obs}}}^2}\right]
    \end{split}
\end{equation}
where $\overline{k}$ is the rate at global equilibrium.
If there is a finite characteristic timescale $t_0$ within which all the unique dynamic activities have occurred, $S_{\text{D}}$ will be saturated when $t_{\text{obs}} \gg t_0$.
Specifically for $t_{\text{obs}}=mt_0$, we divide its history into $m$ pieces, as $K(t_{\text{obs}})=\sum_{i=1}^{i=m} K_i(t_0)$.
The resulting distribution is related to $t_0$, as
\begin{equation}
    p(K, t_{\text{obs}}) \sim \mathcal{N}\left(\overline{k}t_{\text{obs}}, \sigma_t^2=[\sigma_0^2 t_0^{-1}]t_{\text{obs}}\right),
    \label{long time probability}
\end{equation}
where $\sigma_0^2 t_0^{-1}$ is a characteristic constant of the system.
Using Eq.~(\ref{long time probability}), the dynamic entropy can be derived, as
\begin{equation}
    \label{long time entropy}
    S_{\text{D}}(t) = \frac{1}{2}\ln t + \ln \left[\sqrt{2\pi e} \sigma_0 t_0^{-\frac{1}{2}} \right].
\end{equation}
Otherwise, $S(t_{\text{obs}})$ would be in-between the ideal static limit as $\mathcal{O}(0)$ and homogeneous limit as Eq.~(\ref{long time entropy}).
We call this as ``dynamically inhomogeneous''.
At high $P$s, $S_{\text{D}}$s converge to theoretical results of Eq.~(\ref{long time entropy}), as shown in Fig.~\ref{fig3}(d).
The characteristic timescale $t_0$ is decreased towards the active end, consistent with the fact that $s_{\text{ref}}$ is low.
While at low $P$s, $S_{\text{D}}$s increase slowly and are greatly off the $\ln t_{\text{obs}}$ trend, indicating that the system is dynamically inhomogeneous.
When $t_{\text{obs}}\to\infty$, the active end remain the shape, as their $S_{\text{D}}$s only differ from a non-time-dependent term $\ln \left[\sqrt{2\pi e} \sigma_0 t_0^{-\frac{1}{2}} \right]$, and are uniformly lifted by $\frac{1}{2}\ln t_{\text{obs}}$.
In contrast, the inactive end remains flat, and thereby the crossing region witnesses a gradual change.
These confirm the occurrence of a transition between two dynamic states and its high order nature.

\begin{figure}[b]
    \centering
    \includegraphics[width=\linewidth]{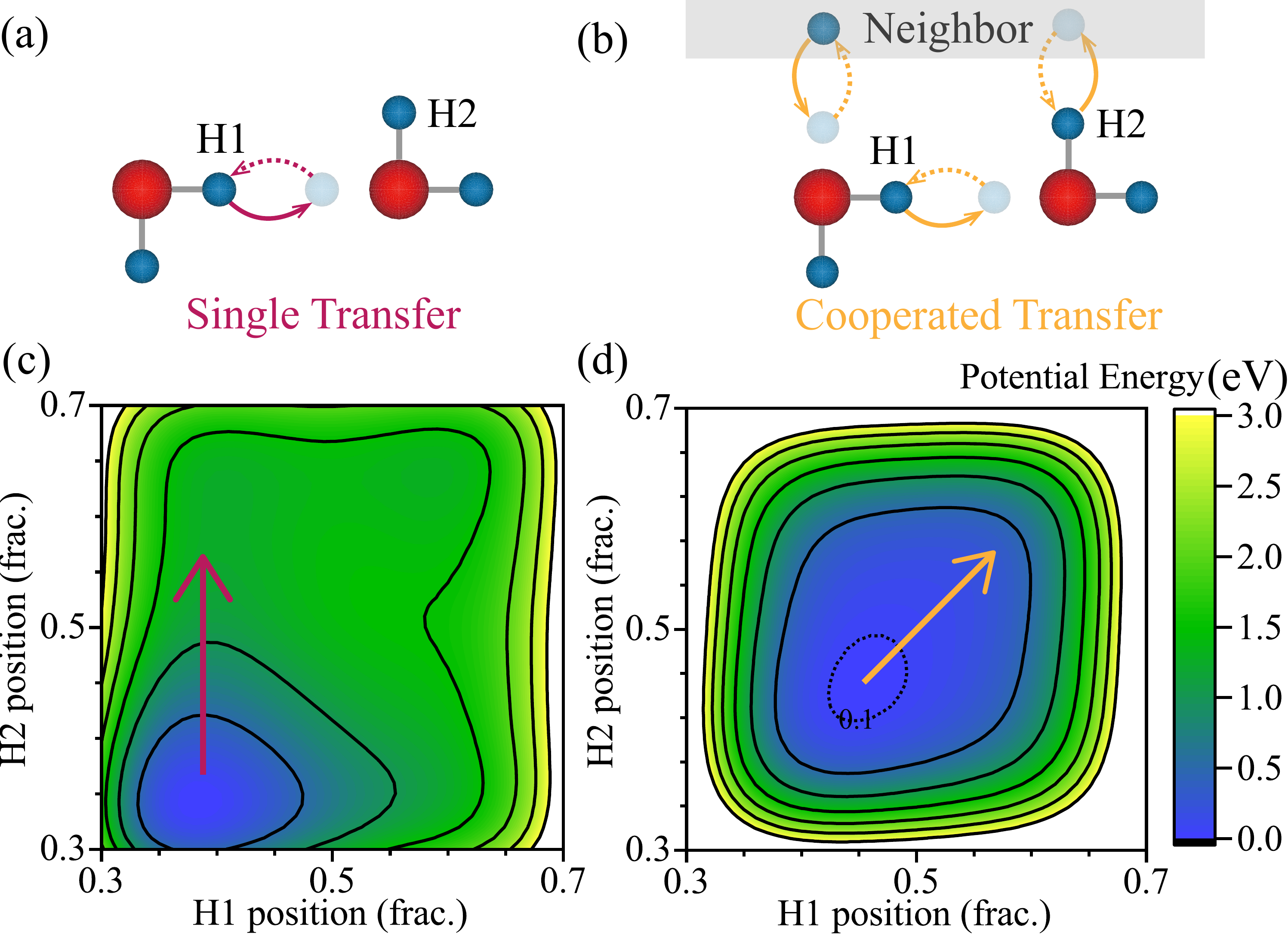}
    \caption{Schematic of the dynamic constraint. When a single transfer occurred (solid line in (a)), there can be two paths to restore the ice rule: (a) the retrieving motion (dashed line) upon strong constraint and (b) the collective motion upon loosen constraint. (c)(d) show the PES upon the transfer of H1 and H2 at 10~GPa and 70~GPa, respectively.}
    \label{fig4}
\end{figure}

Underlying the transition revealed by $s$, the mechanism of proton transfer is detailed via analyzing the PES, where the dynamic constraint is found to be of central importance in inducing different dynamic patterns.
The realistic barrier does not produce absolute confinement, thus the proton is accessible to another component at $t_{\text{obs}}\to\infty$.
When this happens, the system is driven to an energetically uncomfortable state with ice rule temporarily broken (Fig.~\ref{fig4}(a), solid line).
There are two routes to restore the ice rule, either via a retrieving motion of the same proton (Fig.~\ref{fig4}(a), dashed line) or via collective motions involving more protons (Fig.~\ref{fig4}(b), solid line).
When $s_{\text{ref}}$ is large and the dynamic constraint is strong, the system prefers the former.
This can be visualized by the corresponding PES in Fig.~\ref{fig4}(c), as the neighboring protons are not willing to join the transfer from the energetic point of view.
A contrast case is seen when $s_{\text{ref}}$ is small and the dynamic constraint is loosen, the system prefers the latter.
As shown by the flat bottom and collective preference in Fig.~\ref{fig4}(d), the neighboring protons are allowed to participate in a collective transfer~\cite{Ritort2003,Whitelam2004}.
These can also be seen via the coordination number of oxygens shown in the supplemental material~\cite{SI}, as oxygens with 4- and 0- bonded protons which seriously disobey the ice rule only appear at high $P$s.
Structure solely determines dynamics in most cases, naturally establishing a convention that thermodynamic quantities native for phase transitions emerged from equilibrium structures are used to describe transition between different dynamic states.
However, this fails when rare events are of central importance.
Our scheme presents the power of dynamic field in revealing the nature of the transition between dynamic states characterized by different patterns of activities.
Beyond this, the dynamic field offers a numerical approach to access the intrinsic dynamic constraint.
The transition behavior via changing the dynamic field can provide guidance on where different dynamic mechanisms exist for subsequent explorations via experiments or simulations.
It should also be noted that such analysis does not only apply for bcc ice.
Considering the ubiquity of dynamic constraint, we believe this theory would bring new understanding to the fundamental question of the dynamic nature in a wide range of condensed matters.
\begin{acknowledgments}
    The authors are supported by the National Basic Research Programs of China
    under Grand Nos. 2016YFA0300900, the National Science Foundation of China under Grant Nos
    11774003, 11934003, and 11634001.
    XZL thanks Haitao Quan for helpful discussion.
    The computational resources were provided by the supercomputer center in Peking University, China.
\end{acknowledgments}
%

%

\end{document}